\documentclass[10pt,A4paper,fleqn]{amsart}
\usepackage{hyperref}
\usepackage{mathrsfs}
\usepackage{amsfonts}
\usepackage{txfonts}
\usepackage{amsmath}
\usepackage{amssymb}
\usepackage{amsthm}
\usepackage[pdftex]{graphicx}
\usepackage[toc,page,title,titletoc,header]{appendix}
\usepackage{geometry}
\usepackage{upgreek}
\usepackage[T1]{fontenc}
\usepackage[all]{xy}
\usepackage{color}
\usepackage{cancel, ulem}
\usepackage[french,english]{babel}
\usepackage{tikz}

\theoremstyle{plain}
\newtheorem{theo}{Theorem}[section]
\newtheorem*{theo*}{Theorem}

\newtheorem*{lem*}{Lemma}

\theoremstyle{definition}

\theoremstyle{remark}

\newtheorem*{rem*}{Remark}
\newtheorem{rem}{Remark}[section]

\begin{document}

\title{Projective deduction of the non--trivial first integral to the Euler problem: an explicit computation}

\author{Gabriella Pinzari$^{1}$,
Lei Zhao$^{2}$}
\address{$^{1}$ Department of Mathematics, University of Padova, Via Trieste, 63 - 35121 Padova, Italy}
\email{gabriella.pinzari@math.unipd.it}

\address{$^{2}$ School of Mathematical Sciences, Dalian University of Technology, Dalian 116000, China }
\email{{zhao1899@dlut.edu.cn}}


\date\today

\abstract 
The validity of Kepler Laws for the {\it spherical Kepler problem} --- namely, the problem of
the motion of a particle on the unit sphere {in $\mathbb R^3$} undergoing 
an attraction by another particle in the sphere, tangent to the geodesic line between the two and inversely proportional 
to its squared length ---
prompted geometers
to try to interpret such system as a '' projection'' of the familiar Kepler problem in the plane, with the hosting plane given by some affine plane  in $\mathbb R^3$.
At this respect, the most convenient mutual sphere--plane position has been object of a long debate, an account of which can be found in \cite{Albouy2013}.
This fascinating topic, resumed  
by A. Albouy in the aforementioned paper,
has been expanded from the theoretical side in  \cite{Albouy2015}.  Further investigations recently appeared in \cite{AlbouyZhao2019, Zhao1, TakeuchiZhao1, TakeuchiZhao2}.
As remarked in \cite{Albouy2013, Albouy2015}, extensions of the procedure to more dynamical systems  would open to the possibility of finding first integrals to a given dynamical system simply looking at the energy of the mirror problem.
In this note, we focus on the case of
the problem of two fixed centers, already mentioned in \cite{Albouy2013}.
We provide a{n explicit} geometrical  construction allowing  to interpret the first integral of the problem as the energy
of its projection on an ellipsoid.
Compared to previous papers on the same subject,  ours --- besides being based on a somehow different construction --- includes  complete explicit computations. {A byproduct of our construction is the existence of two integrable  mirror  problems (equivalently,   three quadratic integrals, including the energy) for the Kepler problem, which is an aspect of its super-integrability.}
  \endabstract

\maketitle

\section{Introduction}
The { ({\it Euler}--)}{\it two--fixed center problem} is the problem of the motion of a point particle in the Euclidean space $\mathbb R^3$, undergoing Newtonian attraction by two fixed masses. This classical system was studied by {Jacobi},  Euler  and Lagrange \cite{Jacobi, Euler, Lagrange}.
 In Hamiltonian formalism, its 
 Hamilton function is
\begin{eqnarray}\label{J}{\rm J}({\rm p}, {\rm q})=\frac{\|\rm p\|^2}{2}-\frac{m_-}{\|{\rm q}+\rm c\|}-\frac{m_+}{\|{\rm q}-\rm c\|}\,,
\end{eqnarray}
where
\begin{eqnarray}\label{qc}
{\rm c}=(1, 0, 0)\in \mathbb R^3\,,\qquad {\rm q}=(x, y, z)\in {\mathbb R}^3\setminus\{\pm {\rm c}\},
\end{eqnarray}
  $\|\cdot\|$ is  the Euclidean norm in $\mathbb R^3$, and ${\rm p}=\dot {\rm q}=(\dot x, \dot y, \dot z)$,  the ``dot'' denotes derivation with respect to time $t$.  The symplectic form is the canonical one on the phase space.\\
We say that $\rm J$ is {\it integrable} because, as a three--degrees of freedom Hamiltonian, it  exhibits the two following { Poisson-commuting} first integrals, { in addition to}  $\rm J$ itself:
  \begin{eqnarray}\label{Theta E}
  \Theta=({\rm q}\times {\rm p})\cdot {\rm c}\,,\quad 
  {\rm E}=\|{\rm q}\times {\rm p}\|^2+({\rm c}\cdot {\rm p})^2+2 {\rm q}\cdot {\rm c}\left(\frac{m_-}{\|{\rm q}+{\rm c}\|}-\frac{m_+}{\|{\rm q}-{\rm c}\|}\right)\end{eqnarray}
  where $\times$ denotes the skew--product in $\mathbb R^3$.
  Here, by {\it first integral}, we mean a smooth function defined 
  on the whole  phase,
   whose Poisson brackets with $\rm J$ identically vanish, or, equivalently, whose value 
along the orbits of \eqref{J} does not vary.
  While the conservation of the function $\Theta$, component of the angular momentum ${\rm q}\times {\rm p}$ along the direction $\rm c$ of the attracting centers { is }
a mere consequence of the invariance of the Hamiltonian $\rm J$ by rotations about the $\rm c$--axis, the one of $\rm E$ is a remarkable property of the problem{ :}
It is precisely to the derivation of this function as a first integral to $\rm J$  that this note is focused on.
\\
{As outlined in \cite{Albouy2013}, the deduction of $\rm E$ naturally linked to Euler's method involves lenghthy algebraic manipulations, upon which  $\rm E$ comes out more as a  rather cryptic byproduct
	than as the main object of interest.
}
{Of course, this should not sound as a criticism to Euler's work, but as a remark that} finding first integrals in the modern acceptation
recalled above was not a { main} purpose of Euler,
 {just because integrability was only meant   ``by quadratures'' at those times.
Accordingly, his main goal 
was to ``separate'' the degrees of freedom of the ODE associated to the Hamiltonian \eqref{J}.  
 \vskip.1in
 \noindent
Indeed, skipping most details (we refer to, e.g., \cite{Pinzari} and references therein, for more), we recall that Euler's method strongly relies}
on a
set of coordinates which, in the language of Hamiltonian mechanics, consists of
 a canonical, { Mathieu--type transformation} $(A, B, \Theta, \alpha, \beta, \vartheta)\to ({\rm p}, {\rm q})$, where 
\begin{eqnarray*}
\alpha:=\frac{\|{\rm q}+{\rm c}\|+\|{\rm q}-{\rm c}\|}{2}\,,\qquad \beta:=\frac{\|{\rm q}+{\rm c}\|-\|{\rm q}-{\rm c}\|}{2}\,,\qquad \vartheta:=\arg(z, -y)\ {\rm mod}\ 2\pi
\end{eqnarray*}
are the new position coordinates, { and} 
$A$, $B$, $\Theta$ are the generalized impulses respectively  conjugated to $\alpha$, $\beta$, $\vartheta$, with $\Theta$ is precisely as in \eqref{Theta E}. 
The coordinates $(A, B, \Theta, \alpha, \beta, \vartheta)$ are usually referred to as {\it ellipsoidal coordinates}. Once $\rm J$ is expressed through the ellipsoidal coordinates, 
 it turns to be $\vartheta$--independent, {as a consequence of its invariance by rotations about the centers axis}. 
{ T}he main point is that it takes the form
\begin{eqnarray}\label{Jnew}{\rm J}(A, B, \Theta, \alpha, \beta)=\frac{{\rm J}_+(A, \Theta, \alpha)+{\rm J}_-(B, \Theta, \beta)}{\alpha^2-\beta^2}\end{eqnarray}
where ${\rm J}_+(A, \Theta, \alpha)$, ${\rm J}_-(B, \Theta, \beta)$ are suitable functions, which we  omit to specify, depending  only on $(A, \Theta, \alpha)$, $(B, \Theta, \beta)$, respectively.
The form \eqref{Jnew} suggests to proceed by fixing ${\rm J}$ at a given value ${\rm J}_0$ and
switch to the new Hamiltonian $(\alpha^2-\beta^2)({\rm J}(A, B, \Theta, \alpha, \beta)-{\rm J}_0)$,
which is the sum of two  functions
\begin{eqnarray}\label{separate}{\rm J}_+(A, \Theta, \alpha)-{\rm J}_0(\alpha^2-1)\,,\qquad {\rm J}_-(B, \Theta, \beta)-{\rm J}_0(1-\beta^2)\end{eqnarray}
depending separately on $(A, \Theta, \alpha)$, $(B, \Theta, \beta)$. 
 { D}efining a new time { variable} $\tau$  via
\begin{eqnarray*}\frac{d}{d\tau}=(\alpha^2-\beta^2)\frac{d}{dt},
\end{eqnarray*}
one has that the $\tau$--evolution of the couples $(A, \alpha)$, $(B, \beta)$
goes as if they underwent the Hamilton equations
of the functions \eqref{separate},
provided that opposite  values  for the respective energies are chosen,  
 say:
\begin{eqnarray}\label{J+J-}{\rm J}_+(A, \Theta, \alpha)-{\rm J}_0(\alpha^2-1)=-\frac{{\rm E}_0}{2}
\,,\qquad {\rm J}_-(B, \Theta, \beta)-{\rm J}_0(1-\beta^2)=\frac{{\rm E}_0}{2}\,.
\end{eqnarray}
{ Thus rather m}iraculously, an additional constant of motion shows up, namely, the 
{ quantity} ${\rm E}_0$ in \eqref{J+J-}. It is worth underlying that { the} r\^ole of  ${\rm E}_0$ here
is merely  of  splitting the  two--degrees of freedom motion of the Hamiltonian \eqref{Jnew} into separate motions of the
two one--dimensional Hamiltonians \eqref{separate}, under the constraint \eqref{J+J-}. 
By no means ${\rm E}_0$ is to be understood as 
a first integral in the  acceptation recalled above.
However, writing \eqref{Jnew} in the form
\begin{eqnarray}\label{sum}\Big({\rm J}_+(A, \Theta, \alpha)-{\rm J}(A, B, \Theta, \alpha, \beta)(\alpha^2-1)\Big)+\Big({\rm J}_-(B, \Theta, \beta)-{\rm J}(A, B, \Theta, \alpha, \beta)(1-\beta^2)\Big)=0\end{eqnarray}
and naming \begin{eqnarray}\label{E}-\frac{{\rm E}(A, B, \Theta, \alpha, \beta)}{2}\,,\qquad  \frac{{\rm E}(A, B, \Theta, \alpha, \beta)}{2}\end{eqnarray}
the two expressions inside parentheses in \eqref{sum},  
 it turns out that the  function ${\rm E}(A, B, \Theta, \alpha, \beta)$, which is given by \begin{eqnarray}\label{F0}{\rm E}(A, B, \Theta, \alpha, \beta)=
 2\frac{\alpha^2-1}{\alpha^2-\beta^2}{\rm J}_-(B, \Theta, \beta)
 -2\frac{1-\beta^2}{\alpha^2-\beta^2}{\rm J}_+(A, \Theta, \alpha)\,,
 \end{eqnarray} is a first integral { of} ${\rm J}(A, B, \Theta, \alpha, \beta)$. 
 Indeed, comparing \eqref{separate} and \eqref{sum}--\eqref{E}, one has that along  the solutions of  ${\rm J}$ with ${\rm J}(A, B, \Theta, \alpha, \beta)={\rm J}_0$ fixed,
the functions \eqref{E} coincide with the respective left hand sides of \eqref{separate}, hence, ${\rm E}(A, B, \Theta, \alpha, \beta)$ remains constant along the $\tau$--solutions of Hamilton equations of such functions, 
which correspond, by the previous discussion, to  trajectories of $\rm J$ with energy ${\rm J}_0$.  On the other hand, any trajectory of ${\rm J} $ moves on some fixed energy level, hence,  ${\rm E}(A, B, \Theta, \alpha, \beta)$ remains constant on all such trajectories.
 The function  $\rm E$ in \eqref{F0} is precisely the function $\rm E$ in \eqref{Theta E}, after turning back to old coordinates.
\begin{figure}
\centering \includegraphics[width=1\textwidth]{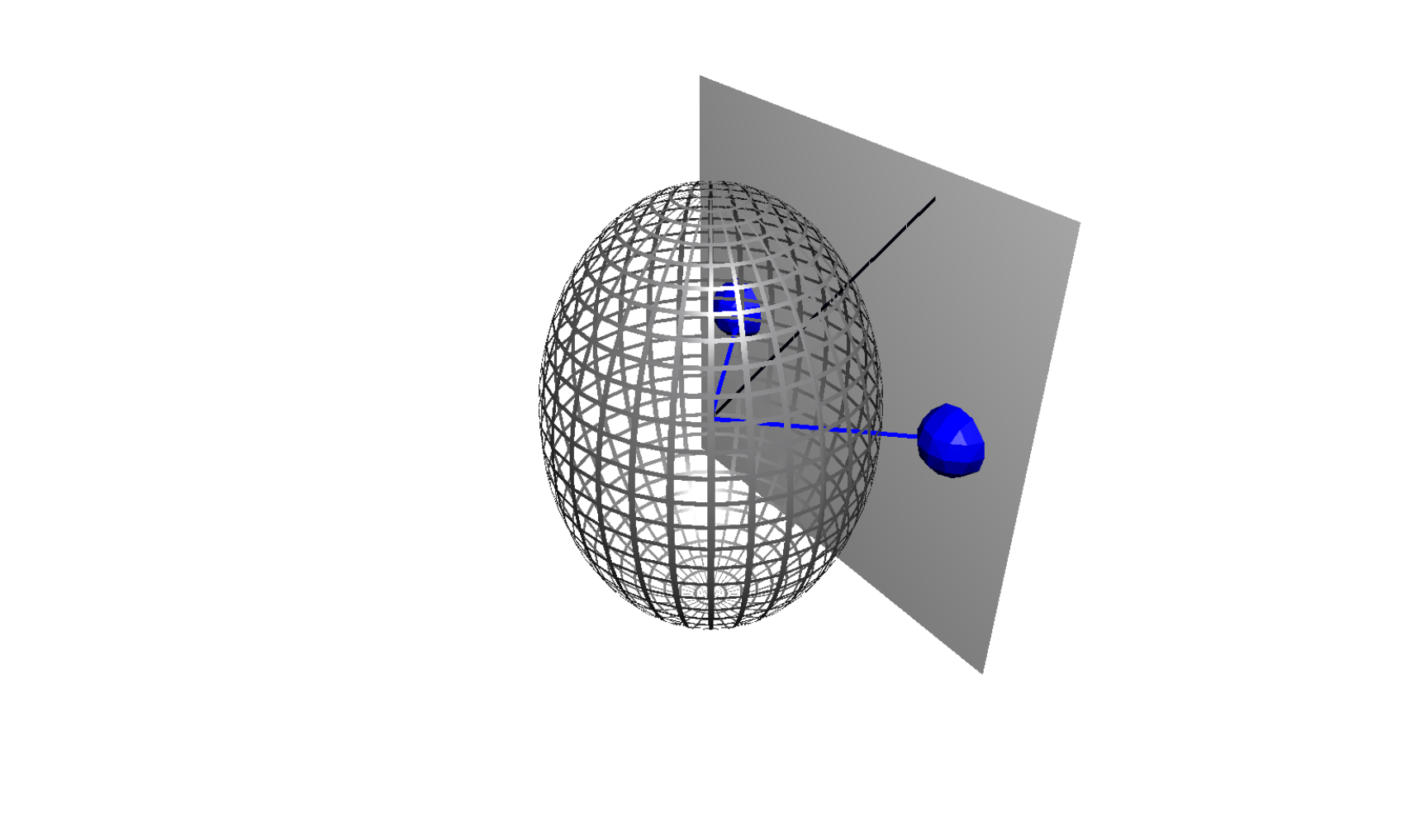} \caption{Plane--Ellipsoid projection}
\label{figure}
\end{figure}
\vskip.1in
\noindent
A
criticism that one can move to the above deduction of $\rm E$ in \eqref{Theta E}
is the rather  computationally involved way it has been obtained through. 
Indeed, a complete proof would require  handling 
the complicated formulae relating $({\rm p}, {\rm q})$ to the ellipsoidal coordinates, which here we have omitted.
Not to mention that the operation we have described of  deducing the function ${\rm E}(A, B, \Theta, \alpha, \beta)$ after finding its level values ${\rm E}_0$'s looks, in some respect, quite artificial
and somehow mysterious.
{To deal with this aspect, in this note we propose a deduction of $\rm E$ strongly inspired to the papers \cite{Albouy2013, Albouy2015}.
{ Indeed, in}  \cite{Albouy2013} an { analysis} of the Kepler problem   (which corresponds to put one of the masses, $m_-$ or $m_+$, to zero in the Hamiltonian \eqref{J}) in the Euclidean space
$\mathbb R^d$ ($d=2$, $3$)
has been proposed
where one looks at the motion of a mirror point on the $d$--dimensional sphere, ${\mathbb S}^d$,  tangent to  $\mathbb R^d$, with the contact point coinciding with the attracting mass position.
It turns out that the motion of such mirror point  shares with the motion of the source  nice properties,  like, for example, having a preserved energy, 
independent of the energy of the source problem,  thus providing 
an independent first integral (in fact, the angular momentum).
}
The argument applies to any {\it SO(2)}--invariant central field.
  Still for the Kepler problem, in \cite{Albouy2015, Zhao1}  a different construction is proposed, with 
   	the attracting particle 
   	slid{ing} 
   	away from the contact point of the sphere with the plane. 
Such a different geometrical implant involves several modifications, among which an affine change of coordinates and the help,  in \cite{Zhao1}, of an additional auxiliary plane, tangent to the sphere from the attracting center. 
	The remarkable byproduct of such modifications is,
	 however, that the energy of the projected point now provides precisely  the Keplerian limit of the function $\rm E$, namely, the right hand side of  \eqref{Theta E} with one of the masses set to zero.
 \\
{It is to be remarked that a word-by-word}
 extension { of the construction in \cite{Zhao1} to the problem  \eqref{J}} 
 does not work in a simple mind fashion because it should involve two different  auxiliary tangent planes.
{The point of view taken up in this note is a variation of the latter, which works for the problem
\eqref{J}. We consider
}
an ellipsoid--plane 
 projection (see Figure \ref{figure}). 
 	Namely, we show that  the projection on a $d$--dimensional ellipsoid of a point $q$ moving on 
	$\mathbb R^d$
	 under the action of the  two--center problem moves as if it were subject to a tangential, conservative vector--field.  Moreover, the corresponding energy results to be a combination of the functions \eqref{J}, \eqref{Theta E}. In other words, we obtain a detailed, simple 
 	derivation of  the integral  $\rm E$, which we believe is a simple way
	to understand the sketches in \cite{Albouy2013}.
When one of the two masses is set to zero, and we are back to the Kepler problem,
 projecting on an ellipsoid rather than on a sphere is the counterpart of the change of norm mentioned 
 in  \cite{Albouy2013}
  and of the affine transformation used in \cite{Zhao1}.
  However, an offshoot of
our construction is that the Kepler problem admits two curved manifolds (namely, the sphere considered in \cite{Albouy2013} and the ellipsoid proposed here) where the  projected motion has a conserved energy. This is another face of its super--integrability (see Rem. \ref{rem: 2.1} below.)
 Further investigations connected to { averaging {and billiard-type problems} in the Euler-two-center  problem} are proposed by the authors in
 \cite{PinzariZhao}.
 
  \section{Projective dynamics of the Euler problem in $\mathbb R^3$}

{Without loss of generality, we fix $d=3$. The case $d=2$ can be obtained fixing the $z$--coordinates to $0$ in all what follows. }
We regard $\mathbb R^3$ as embedded in $\mathbb R^4$, by identifying the former set with the subset ${\mathcal S}:=\mathbb R^3\times\{1\}$ of $\mathbb R^4$. 
We do not set ${\mathcal S}$ as $\mathbb R^3\times\{0\}$, which would adhere to a more usual practice, because the point $0\in \mathbb R^4$ will be reserved as the center of the ellipsoid, which needs not to belong to $\mathcal S$; see below. 
We accordingly extend the ODE associated to \eqref{J} as
\begin{eqnarray}\label{2cp}\ddot q=-m_-\frac{q-c_-}{\|q-c_-\|^3}-m_+\frac{q-c_+}{\|q-c_+\|^3}
\end{eqnarray}
where
\begin{eqnarray*}
c_{\pm}=(\pm \rm c, 1)\in {\mathcal S}\,,\quad q=({\rm q}, 1)\in {\mathcal S}\setminus\{c_\pm\}
\end{eqnarray*}
with $\rm c$, $\rm q$ as in \eqref{qc}. 
As $q-c_\pm=({\rm q}\mp {\rm c}, 0)$,  the three first components of Equation \eqref{2cp} coincide with the original ODE associated to   \eqref{J}, while the fourth one trivializes to ``$0=0$''. 
In the larger space $\mathbb R^4$, we consider the norm (``$*$--norm'' in what follows)
\begin{eqnarray}\label{norm***}\|(x, y, z, w)\|_*:=\sqrt{x^2+\frac{y^2}{2}+\frac{z^2}{2}+w^2}\end{eqnarray}
and define the {codimension--1 manifold}
\begin{eqnarray*}
{\mathcal E}:=\Big\{Q=(X, Y, Z, W)\in \mathbb R^4:\ \|Q\|_*=1\Big\}\,.
\end{eqnarray*}
{In the next, the manifold $\mathcal E$ will be referred to as ``ellipsoid'', as it reduces to an ellipsoid when one of the coordinates $x$, $y$ or $z$ vanishes. As a comparison with \cite{PinzariZhao}, $\mathcal E$ may be interpreted as }
{ 
 the unit-sphere with respect to the $*$--norm.
}
We consider the 
map 
$$\pi:\qquad q=(x, y, z, 1)\in {\mathcal S}\to Q=(X, Y, Z, W)\in {\mathcal E}$$
which assigns to $q\in \mathcal S$ {one} the point{s, and precisely} 
\begin{eqnarray}\label{pi}Q=\pi(q):=\frac{q}{\|q\|_*}
\end{eqnarray}
in $\mathcal E$ on the same ray through the origin of the coordinate system. 
{The map $\pi$ will be also referred to as ``central projection'', as  the origin  of the reference frame is also the symmetry center of $\mathcal E$.}
Note that any $q\in \mathcal S$ has two such points
in $\mathcal E$, given  by $\pm\pi(q)$. The definition in \eqref{pi} corresponds to choose the one such that $W>0$.
Note also  that $\pi$ is well defined for all $q\in \mathcal S$ because $w=\ell(q)=1\ne 0$ implies $\|q\|_*>0$, for all $q\in \mathcal S$.
The inverse map
$$\pi^{-1}:\qquad Q=(X, Y, Z, W)\in {\mathcal E}_+\to q=(x, y, z, 1)\in {\mathcal S}$$
with ${\mathcal E}_+:=\pi({\mathcal S})={\mathcal E}\cap \{W>0\}$
 is given by
\begin{eqnarray}\label{inverse maps1}q=\pi^{-1}(Q):=\frac{Q}{\ell(Q)}\quad {\rm with}\quad \ell(X, Y, Z, W):=W\,.
\end{eqnarray}
Remark finally  that \begin{eqnarray}\label{dual}\ell(Q)\|q\|_*=1\qquad \forall\ q, Q:\ Q=\pi(q)\,.\end{eqnarray} \vskip.1in
\noindent
In this note, we aim to investigate the motion of $Q=\pi(q)\in \mathcal E_+$, provided that $q\in \mathcal S$ undergoes the law \eqref{2cp}.
To this end, we define a new time $\tau$ via \begin{eqnarray}\label{change of time}\frac{d}{d\tau}:=
\frac{1}{\ell(Q(t))^2} \frac{d}{dt}\end{eqnarray}
and let the $\tau$--derivative be denoted with a prime.
In the statement, by {\it tangential vector--field} we mean the vector--field which at   $Q\in \mathcal E_+$ takes the value 
of the  projection of $Q''$
on the plane tangent to $\mathcal E_+$ at $Q$. In principle,
such vector--field should depend on $Q$ and $Q'$. However, this is not the case.
\begin{theo}\label{main}
The tangential vector--field on $ \mathcal E_+$ is $Q'$--independent and, especially, conservative. 
More precisely, the following ``ellipsoidal energy'' 
\begin{eqnarray}\label{energy***}{\rm G}:=\|Q'\|_{*}^2-\sum_{j\in \{\pm\}} \frac{m_j \displaystyle\frac{(c_j, Q)}{\sqrt2}}{
\sqrt{1-\displaystyle\frac{(c_j, Q)^2}{2}}
}\end{eqnarray}
remains constant for all $\tau$.
Turning back to the pre--image $q\in \mathcal S$ of $Q$ and the old time $t$, the function ${\rm G}$ becomes 
\begin{eqnarray}\label{Fnew}
{\rm G}={\rm J}+\frac{\rm E}{2}-\frac{\Theta^2}{4}\end{eqnarray}
with $\rm J$, $\rm E$ and $\Theta$ as in \eqref{J}, \eqref{Theta E}.\end{theo}
\proof 
From \eqref{pi}, we have
\begin{eqnarray*}
	\dot Q=\frac{\dot q \|q\|_{*}- q (Q, \dot q)_{*}}{\|q\|_{*}^2}
\end{eqnarray*}
where $(\cdot, \cdot)_*$ is the inner product associated to the $*$--norm \eqref{norm***}, namely:
\begin{eqnarray}\label{new inner product}
\Big((x_1, y_1, z_1, w_1), (x_2, y_2, z_2, w_2)\Big)_*={x}_1{x}_2+\frac{{y}_1{ y}_2}{2}+\frac{{z}_1{z}_2}{2}+w_1w_2\,.\end{eqnarray}
 With the change of time \eqref{change of time}, ad relation \eqref{dual}, we have
\begin{eqnarray}\label{Qprime}Q'=\frac{\dot Q}{\ell(Q)^2}=\|q\|_{*}^2 \dot Q=\dot q \|q\|_{*}- q (Q, \dot q)_{*}\,.
\end{eqnarray}
Computing another $\tau$--derivative, we get
\begin{eqnarray*}
Q''&=&\|q\|_{*}^2 (\|q\|_{*}^2 \dot Q)^\bullet=\|q\|_{*}^2 \left(\dot q \|q\|_{*}- q (Q, \dot q)_{*}\right)^\bullet\nonumber\\
&=&
\|q\|_{*}^2\left(\ddot q \|q\|_{*}+\dot q 
(Q, \dot q)_{*}
- \dot q (Q, \dot q)_{*}-q 
((Q, \dot q)_{*})^\bullet
\right)
\nonumber\\
&=&\|q\|_{*}^2\left(\ddot q \|q\|_{*}-q 
((Q, \dot q)_{*})^\bullet
\right). 
 \end{eqnarray*}
 The right hand side should now be written using only $Q$, $Q'$, via \eqref{2cp}, \eqref{inverse maps1} and \eqref{Qprime}. The most annoying (even though elementary) part in this task is
 to expand the second term and
 finding the relation which gives $\dot q$, $\ddot q$ 
  as  functions $Q$, $Q'$.  
  However,  for our purposes, it will be sufficient  to write such term as $\widetilde f(Q, Q') Q$, for a suitable $\widetilde f$ which (as we shall see) is not required to know.
We have
\begin{eqnarray}\label{eqQ}
Q''&=&-\frac{\|Q\|^3_{*}}{\ell(Q)^3}\sum_{j\in \{\pm\}} \frac{m_j}{\left\|\frac{Q}{\ell(Q)}-c_j\right\|^3}\left(\frac{Q}{\ell(Q)}-c_j\right)
+\widetilde f(Q, Q') Q\nonumber\\
&=&+\sum_{j\in \{\pm\}} \frac{m_j c_j}{\left\|{Q}-c_j{\ell(Q)}\right\|^3}
+f(Q, Q') Q ,
 \end{eqnarray} 
with $$f(Q, Q'):=-\sum_{j\in \{\pm\}} \frac{m_j}{\ell(Q)\left\|Q-c_j\ell(Q)\right\|^3}
+
\widetilde f(Q, Q')\,.$$
We take the $*$--inner product of \eqref{eqQ} with $Q'$. As $Q\in{\mathcal E}$, we have
 $(Q, Q')_{*}\equiv 0$. Therefore, 
the term $f(Q, Q') Q$ is killed ({ so indeed the exact value of $\widetilde f(Q, Q')$ was irrelevant})
 and we have
\begin{eqnarray}\label{energy}
\frac{1}{2}\left(\|Q'\|_{*}^2\right)'&=&\sum_{j\in \{\pm\}} \frac{m_j (c_j, Q')_{*}}{\left\|{Q}-c_j{\ell(Q)}\right\|^3}. 
\end{eqnarray}
We then compute:
\begin{eqnarray*}(c_-, Q')_{*}&=&\Big((-1, 0, 0, 1), (X', Y', Z', W')\Big)_{*}=-X'+W',\nonumber\\
\left\|{Q}-c_-{\ell(Q)}\right\|&=&\left\|(X, Y, Z, W)-(-1,0,0, 1)W\right\|=\sqrt{(X+W)^2+{Y^2}+{Z^2}}\,.
\end{eqnarray*}
{Using the relation
\begin{eqnarray}\label{ellipsoid}\frac{(-X+W)^2}{2}+\frac{(X+W)^2}{2}+\frac{Y^2}{2}+\frac{Z^2}{2}=X^2+\frac{Y^2}{2}+\frac{Z^2}{2}+W^2=1\end{eqnarray}
we rewrite the latter formula as
\begin{eqnarray*}
	\left\|{Q}-c_-{\ell(Q)}\right\|=\sqrt2\sqrt{1-\frac{(-X+W)^2}{2}}\,.
\end{eqnarray*}
This allows to write the term with $j=-1$ in \eqref{energy} as
\begin{eqnarray*}
 \frac{m_-}{2} \frac{\displaystyle\frac{-X'+W'}{\sqrt2}}{\sqrt{\left(
	1-\displaystyle\frac{(-X+W)^2}{2}
	\right)^3}}=\frac{m_-}{2}\left(
\frac{\displaystyle\frac{-X+W}{\sqrt2}}{\sqrt{
		1-\displaystyle\frac{(-X+W)^2}{2}
		}}
\right)'=\frac{1}{2}
\left(\frac{m_- \displaystyle\frac{(c_-, Q)}{\sqrt2}}{
	\sqrt{1-\displaystyle\frac{(c_-, Q)^2}{2}}
}\right)'\,.
\end{eqnarray*}
Remark that, as the $y$ and $z$--components of $c_-$ vanish, the inner products $(c_-, Q)$ and $(c_-, Q)_*$ coincide.
With similar computations, one has that  the term with $j=+1$ in \eqref{energy}
can be written as 
\begin{eqnarray*}
\frac{1}{2}
	\left(\frac{m_+ \displaystyle\frac{(c_+, Q)}{\sqrt2}}{
		\sqrt{1-\displaystyle\frac{(c_+, Q)^2}{2}}
	}\right)'\,.
\end{eqnarray*}
Then \eqref{energy***} follows.
}
We now check \eqref{Fnew}. Using  the last equality in \eqref{Qprime},  
we have
\begin{eqnarray}\label{last check}\|Q'\|_*^2&=&\Big\|\dot q \|q\|_{*}- q (Q, \dot q)_{*}\Big\|_*^2=\|\dot q_*\|^2\|q\|_{*}^2-2\|q\|_*(Q, \dot q)_{*}(\dot q, q)_*+\|q\|_*^2 (Q, \dot q)_{*}^2\nonumber\\
&=&\|\dot q\|_*^2\|q\|_{*}^2- (q, \dot q)_{*}^2
\end{eqnarray}
where we
have used \eqref{pi}.
To go further, it is convenient to write the $*$--norms and $*$--inner products occurring in\eqref{new inner product} in terms of the Euclidean one:
\begin{eqnarray*}
&&\left\|\left({ x}, { y}, { z}, w\right)\right\|_*=\left\|\left({ x}, \frac{ y}{\sqrt2}, \frac{ z}{\sqrt2}, w\right)\right\|\nonumber\\
&&\left(\left({ x}_1, { y}_1, { z}_1, w_1\right), \left({ x}_2, { y}_2, { z}_2, w_2\right)\right)_*=\left(\left({ x}_1, \frac{{ y}_1}{\sqrt2}, \frac{{ z}_1}{\sqrt 2}, w_1\right), \left({ x}_2, \frac{{ y}_2}{\sqrt2},  \frac{{ z}_2}{\sqrt2}, w_2\right)\right)\,.
\end{eqnarray*}
Then  \eqref{last check} becomes
 \begin{eqnarray}\label{T}\|Q'\|_*^2&=&
\left(\left\|\left({ x}, \frac{ y}{\sqrt2},  \frac{ z}{\sqrt2}\right)\right\|^2+1\right)
\left\|\left(\dot x, \frac{\dot y}{\sqrt2},  \frac{\dot z}{\sqrt2}, 0\right)\right\|^2
-\left(\left({ x}, \frac{ y}{\sqrt2},  \frac{ z}{\sqrt2}, 1\right), \left(\dot x, \frac{\dot y}{\sqrt2},  \frac{\dot z}{\sqrt2}, 0\right)\right)^2
 \nonumber\\
&=&\left\|
\left({ x}, \frac{ y}{\sqrt2},  \frac{ z}{\sqrt2}\right)\times
\left(\dot x, \frac{\dot y}{\sqrt2},  \frac{\dot z}{\sqrt2}\right)
\right\|^2+\left\|\left(\dot x, \frac{\dot y}{\sqrt2},  \frac{\dot z}{\sqrt2}, 0\right)\right\|^2\nonumber\\
&=&%
\dot x^2+\frac{\dot y^2}{2}+\frac{\dot z^2}{2}+\frac{1}{2}\big({ x}\dot y-{ y}\dot x\big)^2
 +\frac{1}{4}\big({ y}\dot z-{ z}\dot y\big)^2
+\frac{1}{2}\big({ z}\dot x-{ x}\dot z\big)^2
\end{eqnarray}
where $\times$ denotes the skew--product in $\mathbb R^3$, and we have used the well--known equality
$$\|a\|^2\|b\|^2-(a, b)^2=\|a\times b\|^2\qquad \forall\ a\,,\ b\in \mathbb R^3\,.$$ 
To complete the computation, it remains to
plug
the  formula \eqref{pi}
into the second term in in \eqref{energy}. This gives  \begin{eqnarray}\label{U}
&&-\sum_{j\in \{\pm\}} \frac{m_j \displaystyle\frac{(c_j, q)}{\sqrt2\|q\|_*}}{
\sqrt{1-\displaystyle\frac{(c_j, q)^2}{2\|q\|_*^2}}
}=
-\sum_{j\in \{\pm\}} \frac{m_j (c_j, q)}{
\sqrt{2\|q\|_*^2-(c_j, q)^2}
}\nonumber\\
&=&\frac{m_-(x-1)}{\sqrt{(x+1)^2+y^2+z^2}}-\frac{m_+(x+1)}{\sqrt{(x-1)^2+y^2+z^2}}
\end{eqnarray}
having used
$$(c_\pm, q)=\pm x+1\,,\qquad 2\|q\|_*^2-(c_\pm, q)^2=2x^2+y^2+z^2-(\pm x+1)^2=(x\mp 1)^2+y^2+z^2\,.$$
Taking the sum of \eqref{T} and \eqref{U}, we arrive at
\begin{eqnarray*}
{\rm G}&=&
{\dot x}^2+\frac{\dot y^2}{2}+\frac{\dot z^2}{2}+\frac{1}{2}\big({ x}\dot y-{ y}\dot x\big)^2
 +\frac{1}{4}\big({ y}\dot z-{ z}\dot y\big)^2
+\frac{1}{2}\big({ z}\dot x-{ x}\dot z\big)^2\nonumber\\
&+&\frac{m_-(x-1)}{\sqrt{(x+1)^2+y^2+z^2}}-\frac{m_+(x+1)}{\sqrt{(x-1)^2+y^2+z^2}}
\end{eqnarray*}
which is immediately seen to be a rewrite of \eqref{Fnew}. $\quad \square${

\rem{\item[(i)]\label{rem: 2.1} 
As a slight generalization (also indicated in \cite{Albouy2013}), one can consider the case when the two attracting masses are located  at a distance $2a$, 
symmetrically with respect to the plane--ellipsoid tangency point.
Theorem \ref{main} still holds, provided that ${\mathcal E}$  is  taken as  \begin{eqnarray}\label{Ea***}{\mathcal E}_a:=\left\{X^2+\frac{Y^2}{1+a^2}+\frac{Z^2}{1+a^2}+W^2=1\right\}
\end{eqnarray}
and the definitions of $\|\cdot\|_*$ and $(\cdot\,,\cdot)_*$ are accordingly changed. The constants in the formula \eqref{Fnew} also need to be adjusted, as well as the formula for $\rm E$, which will be as in \eqref{Theta E}, but with ${\rm c}$ its obvious definition $(a, 0, 0)$. As for the proof, we limit to underline that, instead of relation   \eqref{ellipsoid}, the following couple of ones should be used, implied by \eqref{Ea***}:
\begin{eqnarray*}
	\left.
	\begin{array}{lll}
\displaystyle	\frac{(-aX+W)^2}{1+a^2}+\frac{(X+aW)^2}{1+a^2}+\frac{Y^2}{1+a^2}+\frac{Z^2}{1+a^2}\\\\
	\displaystyle	\frac{(aX+W)^2}{1+a^2}+\frac{(X-aW)^2}{1+a^2}+\frac{Y^2}{1+a^2}+\frac{Z^2}{1+a^2}
		\end{array}
			\right\}=1\,.
	\end{eqnarray*}
	\item[(ii)]
The generalization considered in item (i)  turns to be useful when one wants to look at the Kepler problem as a limiting case   of the two--centers problem. Indeed, the former can be deduced from the latter in two ways: either taking the two centers merging, namely,  taking $a=0$, or letting one of the attracting masses vanish.
In the former case, ${\mathcal E}_a$ reduces to a spheroid and the attracting mass is precisely at the tangency point between this spheroid and  the hyper--plane $\mathcal S$. This is precisely  the case considered in \cite{Albouy2013}, where ${\rm E}$ reduces to the squared angular momentum. If, instead, one chooses to kill one of the  attracting masses, one sees a point moving on an ellipsoid, with an energy given by the formula \eqref{Fnew}, with  $\rm E$ the value   in  \eqref{Theta E} with one only attracting mass. Such function  is precisely the one discussed in \cite{Pinzari1, GJ, Felder}. Thus, the  Kepler problem has at least two manifolds in $\mathbb R^4$  where the motion of the projected point
 has a conserved energy. This 
is another face of its super--integrability (indeed, the respective energies are independent, and provide the angular momentum and the perihelion anomaly).
}
}

\vskip.1in
\noindent
{\bf Acknowledgements:} G. P. acknowledges the ERC project 677793 (2016--2022). L. Z. is supported by the DFG Heisenberg Programme ZH 605/4-1 and the Fundamental Research Funds for the Central Universities of China.

\end{document}